**Sparking Curiosity in Digital System Design Lectures with Take Home Labs**

Şenol Gülgönül

**Electrical and Electronics Engineering Dept., Ostim Technical University, Ankara Turkey**
senol.gulgonul@ostimteknik.edu.tr

**Abstract**

Digital system design lectures are mandatory in the electrical and electronics engineering curriculum. Besides HDL simulators and viewers, FPGA boards are necessary for the real implementation of HDL, which were previously costly for students. With the emergence of low-cost FPGA boards, the use of take-home labs is increasing. The COVID-19 pandemic has further accelerated this process. Traditional lab sessions have limitations, prompting the exploration of take-home lab kits to enhance learning flexibility and engagement. This study aims to evaluate the effectiveness of a low-cost take-home lab kit, consisting of a Tang Nano 9K FPGA board and a Saleae Logic Analyzer, in improving students' practical skills and sparking curiosity in digital system design. The research was conducted in the EEE 303 Digital Design lecture. Students used the Tang Nano 9K FPGA and Saleae Logic Analyzer for a term project involving PWM signal generation. Data was collected through a survey assessing the kit's impact on learning and engagement. Positive Acceptance: 75% of students agreed or strongly agreed that the take-home lab kit was beneficial. Preference for Lab Types: 60% of students preferred classical weekly lab hours over take-home labs. Increased Curiosity: 65% of students conducted additional, unassigned experiments, indicating heightened interest and engagement. The take-home lab kit effectively aids in learning practical aspects of digital system design and stimulates curiosity, though some students prefer traditional lab sessions for group work.

**Keywords:** Take-Home Lab, Digital System Design, FPGA, Engineering Education.

## 1. Introduction

Field Programmable Gate Arrays (FPGAs) are almost mandatory in digital design education due to their practical implementation of Hardware Description Languages (HDL). While simulators like Icarus Verilog and waveform visualization tools such as GTKWave are invaluable for verifying and visualizing code behavior, FPGAs remain essential for observing how the code performs in real-world applications, bridging the gap between simulation and practical implementation [1,2].

Research studies indicate that the use of FPGAs in Digital Systems lectures increases students' understanding of course content. This is supported by the rise in student ratings of their knowledge and confidence. The study concludes that FPGAs significantly improve student engagement and learning outcomes, making them a valuable tool in digital system design education [3]. There are two ways to use FPGA boards: classical laboratories with weekly defined hours or take-home labs. In classical labs, FPGA boards are used by students in groups during weekly lab hours. In the case of take-home labs, students purchase and own the FPGA board, allowing them to perform the required experiments anytime, anywhere.

A variety of FPGA boards are commonly used to provide students with hands-on experience in implementing and testing digital circuits in digital system design lectures. A popular choice is the Xilinx Basys 3, widely adopted due to its user-friendly interfaces, comprehensive documentation, and compatibility with the Vivado Design Suite [4,5]. For low-cost and beginner-friendly options, boards like the GOWIN Tang Nano series and Lattice iCE40-based boards are gaining traction due to their affordability and open-source toolchain support [6,7]. We preferred the Tang Nano 9K for the EEE 303 Digital System Design lectures due to its low price, faster synthesis, availability, and free IDE.

Despite their significant benefits, laboratory experiments have several limitations:

- Laboratory hours are limited due to shared access with other courses, reducing students' flexibility to conduct experiments at their preferred times.
- Limited lab hours do not allow students to fully digest the subject, leading them to rush through experiments and report results.
- Students typically work in groups, which can result in uneven participation, with some students taking the lead while others passively follow.
- During and following the COVID-19 pandemic, students have become hesitant to handle and use shared test equipment [8, 9].
- The high cost of laboratory equipment (e.g., IDL-800A) makes it impractical for personal or home use [10].

Take-home labs enhance self-directed learning and troubleshooting skills more effectively than in-person labs, providing more time for deeper learning [11]. This method also removes infrastructure requirements and relaxes the schedule constraints for both professors and students, making the system more scalable [12]. A survey of students' experiences using The BitBoard at The University of Texas at Arlington (UTA) strongly suggests that take-home lab kits are pedagogically effective [13]. Take-home lab kits also facilitate project-oriented learning through term projects, such as designing UART or SPI modules [14]. In the EEE 303 Digital System Design lecture during the 2024 Fall semester, we preferred to have students generate and control a PWM signal with a frequency related to their student number, ensuring each student had a slightly different project. Currently, laboratories based on real systems are positioned as the main tool for acquiring practical skills, and take-home Labs are adapted to the changes promoted in student-centered educational paradigms [15].

## 2. Materials and Methods

In the EEE 303 Digital Design lecture of Electrical and Electronics Engineering, we follow Morris Mano and Michael D. Ciletti's Digital Design textbook [16]. We teach Verilog Hardware Description Language (HDL), which is easier for students to understand compared to VHDL. To test and verify the developed HDL code, we use the Icarus simulator and GTKWave viewer. As we approach the midpoint of the semester, it has become clear that the simulator is no longer sufficient, and there is a noticeable need among students to transition to physical applications..

After the midterm exam, we began practicing with the Tang Nano 9K board and Saleae Logic Analyzer. Students were also asked to purchase their own take-home lab sets. A term project was assigned, to be completed at any time within one month. The project for the 2024 fall semester involved generating a PWM signal using the FPGA board. The frequency of the PWM signal was set to 1ABC Hz, where ABC represents the last three digits of the student's number. At the end of the lecture, a survey was conducted using Microsoft Forms to measure the effectiveness of the take-home lab experience. A sample Verilog code from a student submission is shown in Fig. 1.

```verilog
module pwm_generator (
    input wire clk,      // Clock signal of the FPGA
    output reg pwm_out   // PWM signal output
);
    parameter CLOCK_FREQ = 27000000; // Tang Nano 9k clock frequency (27 MHz)
    parameter PWM_FREQ = 1019;       // PWM frequency (1ABC -> 1019 Hz)
    parameter DUTY_CYCLE = 50;       // Duty cycle (50%)

    // Total counter value (period)
    localparam integer COUNT_MAX = CLOCK_FREQ / PWM_FREQ;
    // Counter value required for the HIGH state
    localparam integer HIGH_COUNT = (COUNT_MAX * DUTY_CYCLE) / 100;

    reg [31:0] counter = 0; // Counter

    always @(posedge clk) begin
        if (counter < COUNT_MAX - 1)
            counter <= counter + 1; // Increment counter
        else
            counter <= 0;           // Reset counter

        // Generate the PWM signal
        if (counter < HIGH_COUNT)
            pwm_out <= 1; // HIGH state
```

**Figure 1**. An example Verilog code from a student project report

We chose the Tang Nano 9K FPGA due to its low cost (33 USD) compared to the Basys-3 (228 USD), its availability, and the license-free integrated design environment (IDE) GOWIN EDA. In addition to the FPGA kit, we included a Saleae Logic Analyzer clone (8 USD) to visualize and measure digital signals generated by the FPGA. The Saleae Logic Analyzer software, Logic-2, is also provided free of charge. For the take-home digital system design lecture, students set up the Tang Nano 9K and Saleae Logic Analyzer on a breadboard, as shown in Fig. 2.

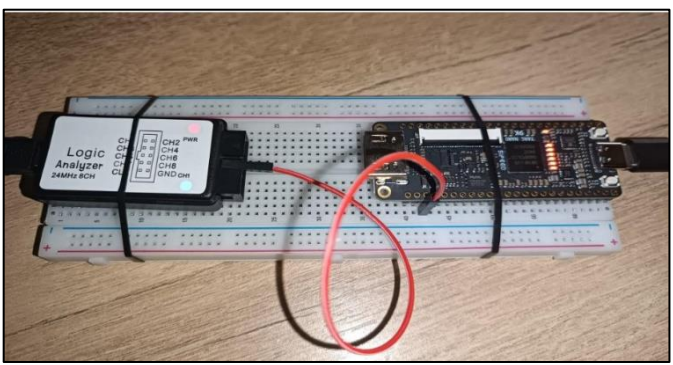

**Figure 2**. An example student setup for *Digital System Design Lecture take-home Lab*

The Tang Nano 9K, powered by Gowin's GW1NR-9 FPGA chip, is a versatile development board featuring 8640 LUT4 logic units, 6480 registers, and 468K bits of block SRAM. It supports multiple display interfaces, including HDMI, RGB, and SPI screen interfaces, and includes a 32Mbit SPI flash for storage. The board is equipped with an onboard 27MHz clock, two PLLs, and six programmable LEDs. Additionally, it offers USB-JTAG and USB-UART debugging capabilities, making it suitable for FPGA verification, RISC-V soft core verification, and functional prototype development [6].

GOWIN EDA is an easy-to-use integrated design environment that offers digital design starting from HDL code to verification. This complete GUI-based environment supports FPGA HDL coding (Verilog, SystemVerilog or VHDL), code synthesis, place and route, bitstream generation, and uploading onto GOWIN FPGAs. It integrates the in-house GowinSynthesis tool for front-end design synthesis and supports both RTL and post-synthesis stages. Users can input RTL files compliant with hardware description languages and constraints files, while post-synthesis input files include netlists generated by user RTL synthesis and required constraints files. Additionally, GOWIN EDA features an IP Core Generator and an online debug tool.GOWIN EDA does not have a simulator thus we have used Icarus and GTKWave for simulation and testing purposes in the lecture.

The Saleae logic analyzer is a compact, highly portable USB-supported device that transforms the notebook into a powerful logic analyzer. This affordable and versatile product is ideal for a wide range of medium-level applications, offering a sampling frequency of up to 24 MHz. It features 8 digital input pins, allowing control of 8 different channels, and provides easy and comfortable usage with 2 ground lines. It is fully compatible with free Saleae Logic software. The analyzer can read and display data from various protocols, including serial UART, I2C, and SPI. An example PWM output at 1.012 kHz from a student submission shown in Fig. 3.

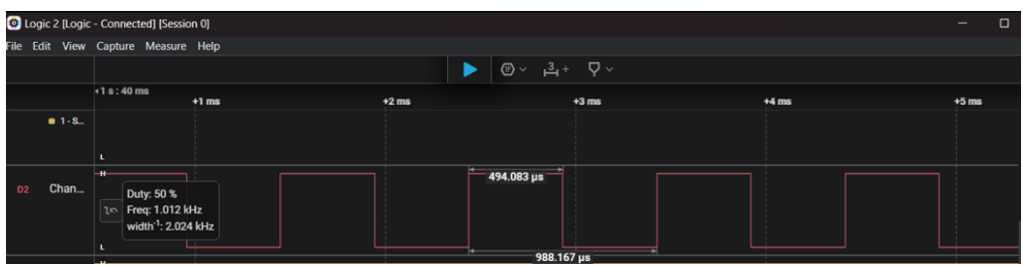

**Figure 3**. A *PWM example from a sudent project report.*

**4. Student Survey Results and Discussion**

Three questions asked to students for evaluation:
1. Please evaluate Tang Nano 9K FPGA board we used in EEE 303 Digital System Design lecture, according to your experience.
2. Do you prefer to learn anytime anywhere by using take-home Lab similar to Tang Nano 9K or classical weekly lab hours?
3. Have you performed any other experiment out of curiosity, that was not asked of you, using Tang Nano 9K FPGA take-home Lab

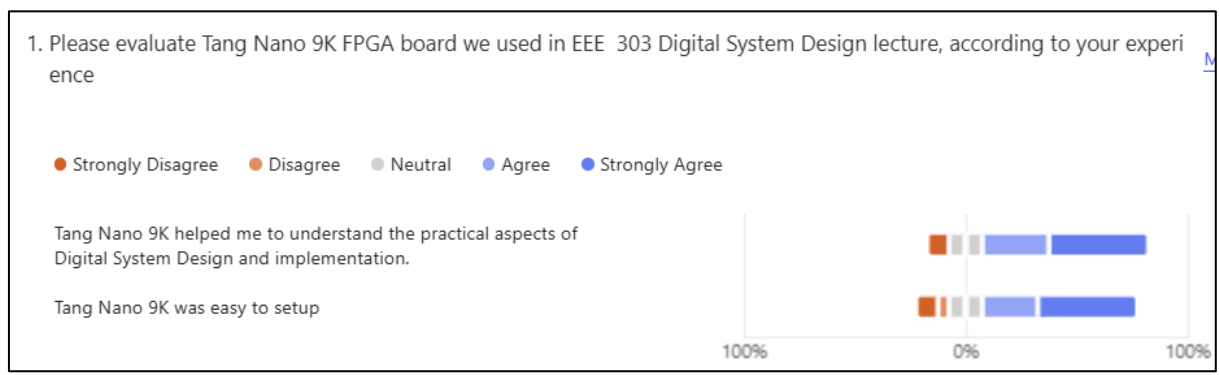

**Figure 4**. *Survey results for the Question 1.*

For the first part of the question, 45% of students 'Strongly Agree', 30% 'Agree', 15% are 'Neutral', and 10% 'Strongly Disagree' as shown in Fig. 4. This indicates a very positive acceptance of the take-home Lab among students. For the second part of the question, student evaluations are 45% 'Strongly Agree', 25% 'Agree', 15% 'Neutral', 5% 'Disagree', and 10% 'Strongly Disagree'. These results suggest that more time is needed to teach the GOWIN EDA and Tang Nano 9K board.

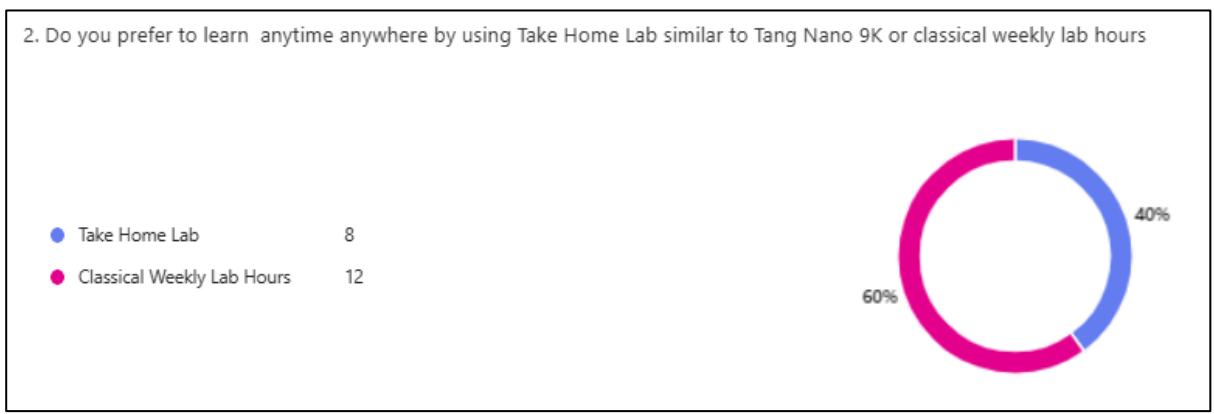

**Figure 5**. *Survey results for the Question 2.*

In the second question, we directly asked students about their preference between the Take-home Lab and Classical Weekly Lab Hours as shown in Fig. 5. Surprisingly, 60% of students preferred the classical weekly lab hours. This result requires further investigation to understand their reasoning. However, it is evident that the Take-home Lab necessitates individual study, whereas in classical labs, students can more easily rely on group work.

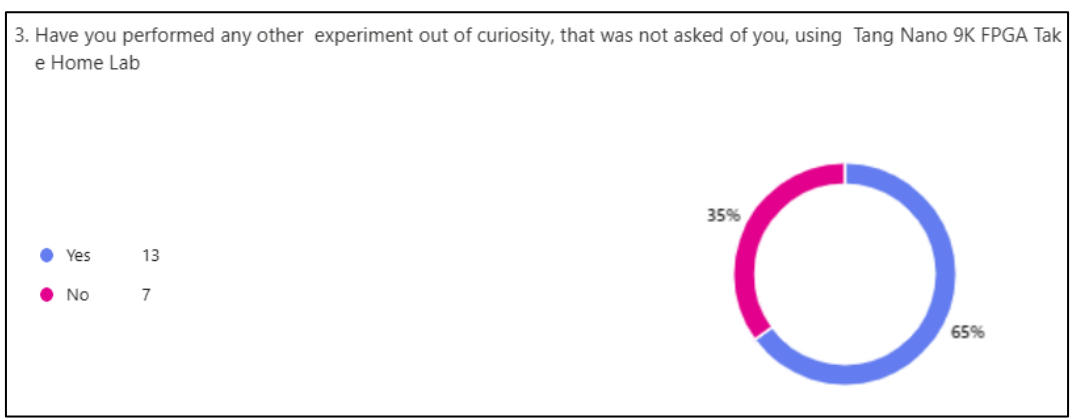

**Figure 6**. *Survey results for the Question 3.*

The third question shown in Fig. 6. aimed to determine if the take-home Lab kit sparked curiosity about digital system design. The results were positive, with 65% of students conducting at least one additional experiment, even though it was not required. This is an intriguing finding, somewhat contradicting the responses to the second question. However, it demonstrates how the Take-home Lab kit stimulates students' curiosity and encourages self-directed learning.

**5. Conclusion**

This study evaluates the educational impact of a take-home Lab kit, composed of a Tang Nano 9K FPGA and a Saleae Logic Analyzer, in a Digital System Design lecture. Survey results indicate that the take-home Lab kit effectively helped students learn the practical aspects of Digital System Design. However, the kit's requirement for individual work was not well-received by most students, who preferred traditional weekly lab sessions. Despite this preference, many students conducted additional, unassigned experiments using the take-home Lab kit. The kit sparked curiosity about the subject and encouraged personal engagement.

**References**

1. https://bleyer.org/icarus/ (Accessed 28.01.2025)
2. https://gtkwave.sourceforge.net/ (Accessed 28.01.2025)
3. Radu, M., & Cole, C., & Dabacan, M. A., & Sexton, S. (2008, June), Extensive Use Of Advanced Fpga Technology In
4. https://digilent.com/reference/programmable-logic/basys-3/start (Accessed 28.01.2025)
5. https://www.amd.com/en/products/software/adaptive-socs-and-fpgas/vivado.html (Accessed 28.01.2025)
6. https://wiki.sipeed.com/hardware/en/tang/Tang-Nano-9K/Nano-9K.html (Accessed 28.01.2025)
7. https://www.latticesemi.com/ice40 (Aceesed 28.01.2025)
8. ROSS, Joel, et al. "Reflections On Engineering Home Lab Kit Use In A Post Pandemic Environment." (2023)
9. Bishop, Z. orcid.org/0000-0002-1894-7433, Howard, T., Lazari, P. et al. (3 more authors)

(2021) Student experiences of practical activities during the COVID-19 pandemic. In: 2021 IEEE Global Engineering Education Conference (EDUCON). EDUCON2021 – IEEE Global Engineering Education Conference, 21-23 Apr 2021, Vienna, Austria

10. https://www.kandh.com.tw/idl-800a-digital-lab-idl-800a.html  (Accessed 28.01.2025)

11. O'Mahony, Tom, et al. "A Take-home Laboratory to Support Teaching Electronics:: Instructors Perspectives and Technical Revisions." Journal on Teaching Engineering 3.1 (2024): 15-29.

12. J. P. Oliver and F. Haim, "Lab at Home: Hardware Kits for a Digital Design Lab," in IEEE Transactions on Education, vol. 52, no. 1, pp. 46-51, Feb. 2009, doi: 10.1109/TE.2008.917191

13. Carroll, Bill D. "The BitBoard-Bridging the Gap from Gates to Gate Arrays." 2016 ASEE Annual Conference & Exposition. 2016.

14. Carroll, Bill, and Jason Losh. "A Novel Project-Oriented System on Chip (SoC) Design Course for Computer and Electrical Engineers." 2022 ASEE Annual Conference & Exposition. 2022.

15. Fernandez, Ricardo M., et al. "New scenarios and trends in non-traditional laboratories from 2000 to 2020." IEEE Transactions on Learning Technologies (2024).

16. Ciletti, Michael D., and M. Morris Mano. Digital design. Hoboken: Prentice-Hall, 2007.